\documentstyle{article}%
\begin{document}
\title{Entropic Approach for Reduction of Amino Acid Alphabets}
\author{Wei-Mou Zheng\\
  {\it Institute of Theoretical Physics, Academia Sinica,
  Beijing 100080, China}}
\date{}
\maketitle

\begin{abstract}
The primitive data for deducing the Miyazawa-Jernigan contact energy or
BLOSUM score metrix is the pair frequency counts. Each amino acid corresponds
to a distribution. Taking the Kullback-Leibler distance of two probability
distributions as resemblance coefficient and relating cluster to mixed
population, we perform cluster analysis of amino acids based on the frequecy
counts data. Furthermore, Ward's clustering is also obtained by adopting
the average score as an objective function. An ordinal cophenetic is
introduced to compare results from different clustering methods.
\end{abstract}

\bigskip

\leftline{\bf Introduction}

Experimental investigation has strongly suggested that protein folding can
be achieved with fewer letters than the 20 naturally occuring amino acids
(Chan, 1999; Plaxco {\it et al.}, 1998). The native structure and
physical properties
of protein Rop is maintained when its 32-residue hydrophobic core is formed
with only Ala and Leu residues (Munson {\it et al.}, 1994). Another example
is the five-letter alphabet of Baker's group for 38 out of 40 selected sites
of SH3 chain (Riddle {\it et al.}, 1997). The mutational tolerance can be
high in many regions of protein sequences. Heterogeneity or diversity in
interaction must be present for polypeptides to have protein-like
properties. However, physics and chemistry for polypeptide chain consisting
of fewer than 20 letters may be sufficiently simplified for a thorough
understanding of the protein folding.

A central task of protein sequence analysis is to uncover the exact nature
of the information encoded in the primary structure. We still cannor read
the language describing the final 3D fold of an active biological
macromolecule. Compared with DNA sequence, protein sequence is generally
much shorter, but the size of the alphabet five times larger. A proper
coarse graining of the 20 amino acids into fewer clusters is important for
improving the signal-to-noise ratio when extracting information by
statistical means.

Based on Miyazawa-Jernigan's (MJ) residue-residue statistical potential
(Miyazawa and Jernigan, 1996), Wang and Wang (1999) (WW) reduced the
alphabet. They introduced a `minimal mismatch' principle to ensure that all
interactions between amino acids belonging to any two given groups are as
similar to one another as possible. The knowledge-based MJ potential is
derived from the frequencies of contacts between different amino acid
residues in a set of known native protein structure database. Murphy,
Wallqvist and Levy (2000) (MWL) approached the same problem using the
BLOSUM metrix derived by Henikoff and Henikoff (1992). The metrix is deduced
from amino acid pair frequencies in aligned blocks of a protein sequence
database, and widely used for sequence alignment and comparison.

The problem
of alphabet reduction may be viewed as cluster analysis, which is a well
developed topic (Romesburg, 1984; Sp\"ath, 1985). WW used the mismatch as an
objective function without any resemblance measure. MWL adopted a cosine-like resemblance coefficient (with a non-standard normalization) from the
BLOSUM score metrix without any objective function, and took the arithmetic
mean of scores to define the cluster center. It is our purpose to propose an
entropic algorithm for selecting reduced alphabet in a consistent and
systematic way.

\leftline{\bf Materials and methods}

Either the MJ contact energies or BLOSUM score metrices are deduced from the
primitive frequency counts of amino acid pairs. Taking the BLOSUM metrix as
an example for specificity, following Henikoff and Henikoff (1992), we
denote the total number of amino acid $i$, $j$ pairs $(1\le i,j \le 20)$ by
$f_{ij}$. It is convenient to introduce another set of $f'_{ij}$ with
$f'_{ij}=f_{ij}/2$ for $i\not =j$ and $f'_{ii}= f_{ii}$, which defines a
joint probability for each $i$, $j$ pair
\begin{equation}
q'_{ij}= f'_{ij}/f, \qquad f= \sum_{i=1}^{20}\sum_{j=1}^{20} f'_{ij}.
\end{equation}
The probability for the amino acid $i$ to occur is then
\begin{equation}
p_i= \sum_{j=1}^{20} q'_{ij}.
\end{equation}
The BLOSUM score corresponds to the logarithm of odds 
\begin{equation}
s_{ij}= \log_2[q'_{ij}/(p_ip_j)].
\end{equation} 
Each Amino acid $i$ may be described by the conditional probability vector
$\{p(j|i)\}_{j=1}^{20}$ with $p(j|i)\equiv q'_{ij}/p_i$. In the language of
cluster analysis, the objects are the 20 amino acids, and the attributes are
$p(j|i)$.

A ruler to measure the similarity between the distributions $\{p_i\}$ and
$\{q_i\}$ is the Kullback-Leibler distance $D$ (also called relative entropy)
of the probability distributions $q$ from $p$ (Kullback, 1959; Kullback
{\it et al.}, 1987; Sakamoto {\it et al}, 1986):
\begin{equation}
D(p,q)= \sum_i p_i \log (p_i/q_i).
\end{equation}
This distance is always non-negative, and not symmetric in general. We may
make symmetrization to use $D=[D(p,q)+D(q,p)]/2$. It will be used as the
resemblace coefficient or distance for clustering. For frequancy counts,
clustering two amino aids is just merging or summing up their counts. A
cluster then corresponds to a mixed population. That is, the cluster center
of amino acid $i$ and $j$ is described by
\begin{equation}
q'_{i\&j,k}= q'_{ik}+q'_{jk}, \qquad p_{i\&j}= p_i+p_j.
\end{equation}
With the resemblance coefficient and cluster center defined, routine cluster
algorithms, such as the centroid method, may be applied.
  
Henikoff and Henikoff (1992) defined the average mutual information or the
average score:
\begin{equation}
H=\sum_{i=1}^{20}\sum_{j=1}^{20} q'_{ij}s_{ij} =\sum_{i=1}^{20}
\sum_{j=1}^{20} q'_{ij} \log_2 \frac{q'_{ij}} {p_ip_j},
\end{equation}
which is again a Kullback-Leibler distance. The difference between $H$ after
and before clustering of $i$ and $j$ is related to terms like
\begin{equation}
(q_{ik}+q_{jk})\log \frac {q_{ik}+q_{jk}}{(p_i+p_j)p_k} - q_{ik}\log \frac
{q_{ik}}{p_ip_k}
-q_{jk}\log \frac {q_{jk}}{p_jp_k},
\end{equation}
which, by introducing $x_i=q_{ik}/p_i$, $x_j=q_{jk}/p_j$, $\omega_i=
p_i/(p_i+p_j)$, $\omega_j =p_j/(p_i+p_k)$ and $\langle f(x)\rangle = \omega_i
f(x_i) +\omega_j f(x_j)$, is proportional to $f(\langle x\rangle ) -\langle
f(x)\rangle$ with $f(x) =x\log x$. 
From the Jessen theorem for convex function ($x\log x$ here) (Rassias, 2000;
Rassias and Srivastava, 1999), $H$ never increases after each step of
clustering. To make the average score as closer to that before a
coarse-graining as possible, we should maximize $H$. This average mutual
information $H$ can be chosen as the objective function for clustering with
respect to scores. Compared with the above approach based on the conditional
probability $p(j|i)$, this objective function also takes abundance of amino
acids into account. We shall use Ward's methord (Romesburg, 1984) to perform
clustering.

\leftline{\bf Results}

By means of the entropic Kullback-Leibler distance, defining the center of
cluster by the distribution of the mixed population, we conduct cluster
analysis on the MJ frequency counts with the centroid method. The result of
the hierachical steps of clustering is shown in Table I. This will be
refered to as the MJ-clustering. We do see Baker's five representative
letters (AIGEK) at step 14, which ends at 6 clusters including the cluster
consisting of the extraordinary sigle member Cys.

Our most cluster analysis is done based on the BLOSUM 62 frequency counts.
The counterpart of Table I for BLOSUM is Table II. Taking the average score
$H$ as the objective function for maximization, the clustering result of
Ward's method is given in Table III. These two clusterings will be referred
to as the HH- and BL-clustering, respectively. For the BL-clustering, when
number of clusters becomes smaller, the average score decreases faster as
shown in Fig.~1. When the total number of clusters is three, the score drops
to about the half of its original value.

Clustering result can be represented by a tree. The cophenetic metrix built
by tracing distances along the tree is equivalent to the tree. The
correlation between
the cophenetic matrix and resemblance matrix is often used to measure the
quality of clustering. We introduce the ordinal cophenetic metrix by taking
the clustering depth as the distance. For example, Y and Q group together
at step 5 in Table I. The YQ element of the ordinal cophenetic metrix is 5
as shown in Table IV, where the lower and upper matrices correspond to
Tables I and II, respectively. In this way we ignore some numerical
details, and focus on the order of the nodes in the tree. We compare the
BL-clustering with the MJ- and HH-clusterings by calculating the difference
between their ordinal cophenetic matrices. As shown in Table 5, the two
clusterings HH and BL are closer to each other than MJ and BL are. Large
positive and negative values of entries indicate main differences. The
MJ-clustering prefers F to group with M, and Q with Y, while the BL-or
HH-clustering prefers F to group with Y, and Q with E. In all the three
clusterings the separation of hydrophobic and hydrophilic groups is rather
clear.

\leftline{\bf Discussion}

We have done cluster analysis also based on the BLOSUM 50 and 90. The
results are very close to those obtained for the BLOSUM 62.

The clustering based on MJ shows avident discrepency from that based on
BLOSUM. From the way obtaining the frequency acounts, the BLOSUM data is
more relavent to evolutional difference of residues, while the MJ data to
structure difference. There are many amino acid difference formulas
(Grantham, 1974). From composition $c$ (defined as the atomic weight ratio
of noncarbon elements in end groups to carbons in the side chain), polarity
$p$ and volume $v$ Grantham (1974) derived an amino acid defference matrix,
which exhibits stronger correlation with evolution than the method of
minimum base changes between codons. This difference metrix is also a good
candidate of resemblance metrix for clustering. To least disturb the data,
we perform the UPGMA (unweighted pair-group method using arithmetic average)
clustering (referred to as GR) on the data. The difference in the ordinal
cophenetic metrices is shown in Table VI. Compared with MJ, BL is close to
GR derived from physicochemical properties of amino acids. The average
absolute difference of 190 entries are 1.84 and 2.84 for BL $-$ GR
and HH $-$ GR, respectively. Since different structure regularities prefer
certain residues, residue clustering should not be identical in all
structure subclasses. Structure subclass specific clustering would give us
more insight.

For the BLOSUM data, the direct use of pair frequency counts provides us a
consistent way to derive coarse-grained scores from mixed population. We
think for the MJ data the natural objective function should be the average
contact energy, which is the counterpart of the BLOSUM $H$ by replacing the
logarithm of odds $s_{ij}$ with the contact energy $e_{ij}$. The
coarse-grained contact energy can be deduced from mixed population. This
will furnishe us a consistent way for cluster analysis.

\begin{quotation}
{ This work was finished during the author's visit to the Bartol Research
Institute, University of Delaware. The author thanks Dr.~S.T. Chui for the
warm hospitality and fruitful discussions. This work was supported in part
by the Special Funds for
Major National Basic Research Projects, the National Natural Science
Foundation of China and Research Project 248 of Beijing.}
\end{quotation}

\leftline{\bf References}
{\parskip=0pt \parindent=0pt
Chan,H.S. (1999) {\it Nature Struct. Biol.}, {\bf 6}, 994--996.

Grantham,R. (1974) {\it Science}, {\bf 185}, 862--864.

Henikoff,S. and Henikoff,J.G. (1992) {\it Proc. Natl. Acad. Sci. USA},
{\bf 89}, 10915--10919.

Kullback,S., Keegel,J.C. and Kullback,J.H. (1959) {\it Information Theory
and Statistics}, Wiley, New York.

Kullback,S. (1987) {\it Topics in Statistical Information Theory}, Springer,
Berlin.

Miyazawa,S. and Jernigan,R.L. (1996) {\it J. Mol. Biol.}, {\bf 256}, 623--644.

Munson,M., O'Brien,R., Sturtevant,J.M. and Regan,L. (1994) {\it Protein
Sci.}, {\bf 3}, 2015--2022.

Murphy,L.R., Wallqvist,A. and Levy,R.M. (2000) {\it Protein Eng.}, {\bf 3},
149--152.

Plaxco,K.W., Riddle,D.S., Grantcharova,V.P. and Baker,D. (1998) {\it Curr.
Opin. Struct. Biol.}, {\bf 8}, 80--85.

Rassias,T.M. (eds) (2000) {\it Survey on Classical Inequalities}, Kluwer
Academic, Dordrecht.

Rassias,T.M. and Srivastava,H.M. (eds) (1999) {\it Analytic and Geometric
Inequalities and Applications}, Kluwer Academic, Dordrecht.

Riddle,D.S., Santiago,J.V., Bray-Hall,S.T., Doshi,N., Grantcharova,V.P.,
Yi,Q. and Baker,D. (1997) {\it Nature Struct. Biol.}, {\bf 4}, 805--809.

Romesburg,H.C. (1984) {\it Cluster Analysis for Researchers}, Lifetime
Learning Publications, Belmont.

Sakamoto,T., Ishiguro,M. and Kitagawa,G. (1986) {\it Akaike Information
Criterion Statistics}, KTK Scientific, Tokyo.

Sp\"ath,H. (1985) {\it Cluster Dissection and Analysis: Theory, FORTRAN
Program, Examples}, Ellis Horwood, New York.

Wang,J. and Wang,W. (1999) {\it Nature Struct. Biol.}, {\bf 6}, 1033--1038.
}
\bigskip

Table I. Clustering based on the MJ pair frequency counts with the centroid
method. The first column indicates the step in the hierachical clustering.

{\baselineskip=0pt
\begin{verbatim}
 0  C A V I L M F Y Q P W N S T G H D E R K 
 1  C A V IL  M F Y Q P W N S T G H D E R K 
 2  C A V IL  MF  Y Q P W N S T G H D E R K 
 3  C A VIL   MF  Y Q P W N S T G H D E R K 
 4  C A VIL   MF  Y Q P W N ST  G H D E R K 
 5  C A VIL   MF  YQ  P W N ST  G H D E R K 
 6  C A VIL   MF  YQP   W N ST  G H D E R K 
 7  C A VIL   MF  YQP   W N ST  G H D E RK 
 8  C A VILMF     YQP   W N ST  G H D E RK 
 9  C A VILMF     YQP   W NST   G H D E RK 
10  C A VILMF     YQPW    NST   G H D E RK 
11  C A VILMF     YQPW    NSTG    H D E RK 
12  C A VILMF     YQPW    NSTG    H DE  RK 
13  C A VILMF     YQPWNSTG        H DE  RK 
14  C A VILMF     YQPWNSTGH         DE  RK 
15  C AVILMF      YQPWNSTGH         DE  RK 
16  CAVILMF       YQPWNSTGH         DE  RK 
17  CAVILMFYQPWNSTGH                DE  RK 
18  CAVILMFYQPWNSTGHDE                  RK 
19  CAVILMFYQPWNSTGHDERK 
\end{verbatim} }

Table II. Clustering based on the BLOSUM62 frequency counts with the
centroid method.
{\baselineskip=0pt
\begin{verbatim}
 0  W F Y L M I V H N D R K Q E S T A G P C
 1  W F Y LM  I V H N D R K Q E S T A G P C
 2  W F Y LM  IV  H N D R K Q E S T A G P C
 3  W F Y LM  IV  H N D RK  Q E S T A G P C
 4  W F Y LMIV    H N D RK  Q E S T A G P C
 5  W F Y LMIV    H N D RK  QE  S T A G P C
 6  W F Y LMIV    H N D RK  QE  ST  A G P C
 7  W F Y LMIV    H N D RK  QE  STA   G P C
 8  W FY  LMIV    H N D RK  QE  STA   G P C
 9  W FY  LMIV    H N D RKQE    STA   G P C
10  W FY  LMIV    H ND  RKQE    STA   G P C
11  W FY  LMIV    H NDRKQE      STA   G P C
12  W FY  LMIV    H NDRKQESTA         G P C
13  W FY  LMIV    HNDRKQESTA          G P C
14  W FY  LMIV    HNDRKQESTAG           P C
15  W FYLMIV      HNDRKQESTAG           P C
16  W FYLMIV      HNDRKQESTAGP            C
17  W FYLMIVHNDRKQESTAGP                  C
18  W FYLMIVHNDRKQESTAGPC
19  WFYLMIVHNDRKQESTAGPC
\end{verbatim}}

\newpage
Table III. Clustering based on the BLOSUM62 frequency counts with Ward's
method.
{\baselineskip=0pt
\begin{verbatim}
 0  W F Y L M I V H N D R K Q E S T A G P C
 1  W F Y LM  I V H N D R K Q E S T A G P C
 2  W F Y LM  I V H N D RK  Q E S T A G P C
 3  W F Y LM  I V H N D RK  QE  S T A G P C
 4  W F Y LM  IV  H N D RK  QE  S T A G P C
 5  W FY  LM  IV  H N D RK  QE  S T A G P C
 6  W FY  LM  IV  H N D RK  QE  ST  A G P C
 7  W FY  LM  IV  H N D RK  QE  STA   G P C
 8  W FY  LMIV    H N D RK  QE  STA   G P C
 9  W FY  LMIV    H ND  RK  QE  STA   G P C
10  W FY  LMIV    H ND  RKQE    STA   G P C
11  WFY   LMIV    H ND  RKQE    STA   G P C
12  WFY   LMIV    HND   RKQE    STA   G P C
13  WFY   LMIV    HND   RKQE    STA   G PC
14  WFY   LMIV    HNDRKQE       STA   G PC
15  WFY   LMIV    HNDRKQE       STAG    PC
16  WFY   LMIV    HNDRKQE       STAGPC
17  WFYLMIV       HNDRKQE       STAGPC
18  WFYLMIV       HNDRKQESTAGPC
19  WFYLMIVHNDRKQESTAGPC
\end{verbatim}
}
Table IV. Ordinal cophenetic matrices of the MJ-clustering (lower) and
HH-clustering (upper).
{\baselineskip=0pt\small
\begin{verbatim}
   W  F  Y  L  M  I  V  H  N  D  R  K  Q  E  S  T  A  G  P  C
W    19 19 19 19 19 19 19 19 19 19 19 19 19 19 19 19 19 19 19
F 17     8 15 15 15 15 17 17 17 17 17 17 17 17 17 17 17 17 18
Y 10 17    15 15 15 15 17 17 17 17 17 17 17 17 17 17 17 17 18
L 17  8 17     1  4  4 17 17 17 17 17 17 17 17 17 17 17 17 18
M 17  2 17  8     4  4 17 17 17 17 17 17 17 17 17 17 17 17 18
I 17  8 17  1  8     2 17 17 17 17 17 17 17 17 17 17 17 17 18
V 17  8 17  3  8  3    17 17 17 17 17 17 17 17 17 17 17 17 18
H 14 17 14 17 17 17 17    13 13 13 13 13 13 13 13 13 14 16 18
N 13 17 13 17 17 17 17 14    10 11 11 11 11 12 12 12 14 16 18
D 18 18 18 18 18 18 18 18 18    11 11 11 11 12 12 12 14 16 18
R 19 19 19 19 19 19 19 19 19 19     3  9  9 12 12 12 14 16 18
K 19 19 19 19 19 19 19 19 19 19  7     9  9 12 12 12 14 16 18
Q 10 17  5 17 17 17 17 14 13 18 19 19     5 12 12 12 14 16 18 
E 18 18 18 18 18 18 18 18 18 12 19 19 18    12 12 12 14 16 18
S 13 17 13 17 17 17 17 14  9 18 19 19 13 18     6  7 14 16 18
T 13 17 13 17 17 17 17 14  9 18 19 19 13 18  4     8 14 16 18
A 17 15 17 15 15 15 15 17 17 18 19 19 17 18 17 17    14 16 18
G 13 17 13 17 17 17 17 14 11 18 19 19 13 18 11 11 17    16 18
P 10 17  6 17 17 17 17 14 13 18 19 19  6 18 13 13 17 13    18
C 17 16 17 16 16 16 16 17 17 18 19 19 17 18 17 17 16 17 17
\end{verbatim}}

\newpage
Table V. Difference of the BL ordinal cophenetic metrix from those of the
MJ- (lower) and HH-clustering (upper).
{\baselineskip=0pt\small
\begin{verbatim}
   W  F  Y  L  M  I  V  H  N  D  R  K  Q  E  S  T  A  G  P  C
W    -8 -8 -2 -2 -2 -2  0  0  0  0  0  0  0  0  0  0  0  0  0
F -6    -3  2  2  2  2  2  2  2  2  2  2  2  2  2  2  2  2  1
Y  1-12     2  2  2  2  2  2  2  2  2  2  2  2  2  2  2  2  1
L  0  9  0     0  4  4  2  2  2  2  2  2  2  2  2  2  2  2  1
M  0 15  0 -7     4  4  2  2  2  2  2  2  2  2  2  2  2  2  1
I  0  9  0  7  0     2  2  2  2  2  2  2  2  2  2  2  2  2  1
V  0  9  0  5  0  1     2  2  2  2  2  2  2  2  2  2  2  2  1
H  5  2  5  2  2  2  2    -1  0  1  1  1  1  5  5  5  4  2  0
N  6  2  6  2  2  2  2 -2    -1  3  3  3  3  6  6  6  4  2  0
D  1  1  1  1  1  1  1 -5 -9     3  3  3  3  6  6  6  4  2  0
R  0  0  0  0  0  0  0 -5 -5 -5    -1  1  1  6  6  6  4  2  0
K  0  0  0  0  0  0  0 -5 -5 -5 -5     1  1  6  6  6  4  2  0
Q  9  2 14  2  2  2  2  0  1 -4 -9 -9    -2  6  6  6  4  2  0 
E  1  1  1  1  1  1  1 -4 -4  2 -9 -9-15     6  6  6  4  2  0
S  6  2  6  2  2  2  2  4  9  0 -1 -1  5  0     0  0  1  0 -2
T  6  2  6  2  2  2  2  4  9  0 -1 -1  5  0  2    -1  1  0 -2
A  2  4  2  4  4  4  4  1  1  0 -1 -1  1  0-10-10     1  0 -2
G  6  2  6  2  2  2  2  4  7  0 -1 -1  5  0  4  4 -2     0 -2
P  9  2 13  2  2  2  2  4  5  0 -1 -1 12  0  3  3 -1  3    -5
C  2  3  2  3  3  3  3  1  1  0 -1 -1  1  0 -1 -1  0 -1 -4
\end{verbatim}}

Table VI. Difference of the MJ- (lower) and BL-clustering (upper) ordinal
cophenetic metrices from that of the GR-clustering.
{\baselineskip=0pt\small
\begin{verbatim}
   W  F  Y  L  M  I  V  H  N  D  R  K  Q  E  S  T  A  G  P  C
W    -4 -4  2  2  1  1  1  1  1  1  1  1  1  1  1  1  1  1  0
F  2     2  7  7  7  7  1  1  1  1  1  1  1  1  1  1  1  1  0
Y -5 14     7  7  7  7  1  1  1  1  1  1  1  1  1  1  1  1  0
L  2 -2  7    -1  0  0  1  1  1  1  1  1  1  1  1  1  1  1  0
M  2 -8  7  6     6  0  1  1  1  1  1  1  1  1  1  1  1  1  0
I  2 -2  7  0  6    -4  1  1  1  1  1  1  1  1  1  1  1  1  0
V  2 -2  7 -5  0 -5     1  1  1  1  1  1  1  1  1  1  1  1  0
H -4 -1 -4 -1 -1 -1 -1    -4 -3  3  3  9  5  2  1  1  1  1 -1
N -5 -1 -5 -1 -1 -1 -1 -2     5 -2 -2 -2 -2  4  1  1  1  1 -1
D  0  0  0  0  0  0  0  2 14    -2 -2 -2 -2  4  1  1  1  1 -1
R  1  1  1  1  1  1  1  8  3  3    -4 -1 -1  2  1  1  1  1 -1
K  1  1  1  1  1  1  1  8  3  3  1    -1 -1  2  1  1  1  1 -1
Q -8 -1-13 -1 -1 -1 -1  9 -3  2  8  8    -6  2  1  1  1  1 -1 
E  0  0  0  0  0  0  0  9  2 -4  8  8  9     2  1  1  1  1 -1
S -5 -1 -5 -1 -1 -1 -1 -2 -5  4  3  3 -3  2   -11-10 -2 -1 -3
T -5 -1 -5 -1 -1 -1 -1 -3 -8  1  2  2 -4  1-13    -5  2  4 -3
A -1 -3 -1 -3 -3 -3 -3  0  0  1  2  2  0  1  0  5     2  9 -3
G -5 -1 -5 -1 -1 -1 -1 -3 -6  1  2  2 -4  1 -6 -2  4     3 -3
P -8 -1-12 -1 -1 -1 -1 -3 -4  1  2  2-11  1 -4  1 10  0    -6
C -2 -3 -2 -3 -3 -3 -3 -2 -2 -1  0  0 -2 -1 -2 -2 -3 -2 -2
\end{verbatim}}

Fig.~1 Relationship between everage score and number of clusters. (Here the
score is in the natural logarithm instead of taking base 2.) 

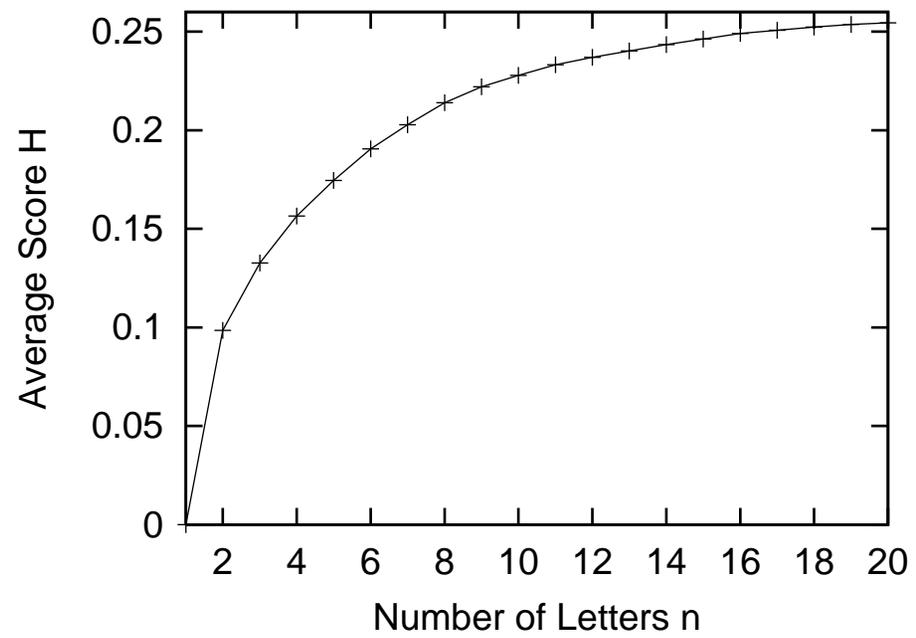
\end{document}